\documentclass[showpacs,preprintnumbers,amsmath,amssymb,prl,twocolumn]{revtex4}
\usepackage{amsfonts}
\usepackage{mathrsfs}
\usepackage{amsmath}
\usepackage{graphicx}
\usepackage{dcolumn}
\usepackage{bm}
\usepackage{verbatim}
\usepackage[T1]{fontenc}
\usepackage{epsfig}

\begin{document}
\title{ Quantum phases with differing computational power }
\author{Jian Cui$^{1,2}$}
\email{cuijian@iphy.ac.cn;}
\author{ Mile Gu$^{2}$}
\email{ceptryn@gmail.com }
\author{  Leong Chuan Kwek$^{2,3}$ }
\author{ Marcelo Fran\c{c}a Santos$^{4}$}
\author{Heng Fan$^{1}$}
\author{Vlatko Vedral$^{2,5,6}$}
 \affiliation{$^1$
Institute of Physics, Chinese Academy of Sciences, Beijing 100190,
China} \affiliation{$^2$ Centre for Quantum Technologies, National
University of Singapore, 3 Science Drive 2, Singapore 117543}
\affiliation{$^3$ National Institute of Education and Institute of
Advanced Studies, Nanyang Technological University, 1 Nanyang Walk,
Singapore 637616}
 \affiliation{$^4$ Departamento de F\'{\i}sica, Universidade Federal de Minas Gerais,
Belo Horizonte, Caixa Postal 702, 30123-970, MG, Brazil}
  \affiliation{$^5$ Atomic and Laser Physics, Clarendon Laboratory,
University of Oxford, Parks Road, Oxford OX13PU, United Kingdom}
\affiliation{$^6$ Department of Physics, National University of
Singapore, Republic of Singapore}
\date{\today}

\begin{abstract}
The observation that concepts from quantum information has generated
many alternative indicators of quantum phase transitions hints that
quantum phase transitions possess operational significance with
respect to the processing of quantum information. Yet, studies on
whether such transitions lead to quantum phases that differ in their
capacity to process information remain limited. Here We show that
there exist quantum phase transitions that cause a distinct
qualitative change in our ability to simulate certain quantum
systems under perturbation of an external field by local operations
and classical communication. In particular, we show that in certain
quantum phases of the $XY$ model, adiabatic perturbations of the
external magnetic field can be simulated by local spin operations,
whereas the resulting effect within other phases results in coherent
non-local interactions. We discuss the potential implications to
adiabatic quantum computation, where a computational advantage
exists only when adiabatic perturbation results in coherent
multi-body interactions.
\end{abstract}
\maketitle

\section{Introduction}
 The study of quantum phase transitions has greatly benefited from
developments in quantum information theory
\cite{REVIEWSOFMODERNPHYSICS,numericalmethods}. We know, for
example, that the extremum points of entanglement and other related
correlations coincide with phase transition points
\cite{nielsenIsing,Osterlohnature,VidalKitaevprl,jiancui,Kit}, and
that different phases may feature differing fidelity between
neighboring states
\cite{fidelity1,fidelity2,fidelity3,fidelity4,fidelity5,fidelity6}.
These observations have helped pioneer many alternative indicators
of phase transitions, allowing the tools of quantum information
science to be harnessed in the analysis of quantum many body
systems\cite{REVIEWSOFMODERNPHYSICS,numericalmethods}. The reverse,
however, remains understudied. If the concepts of quantum
information processing have such relevance to the study of quantum
phase transitions, one would expect that systems undergoing quantum
phase transition would also exhibit different operational properties
from the perspective of information processing. Yet, there remains
little insight on how such relations are apply to quantum
information and computation.

In this paper, we demonstrate using the XY model that different
quantum phases have distinct operational significance with respect
to quantum information processing. We reveal that the differential
local convertibility of ground states undergoes distinct qualitative
change at points of phase transition. By differential local
convertibility of ground states, we refer to the following (Fig. 1):
A given physical system with an adjustable external parameter $g$ is
partitioned into two parties, Alice and Bob. Each party is limited
to local operations on their subsystems (which we call $A$ and $B$)
and classical inter-party communication, i.e., local operations and
classical communication (LOCC). The question is: \textit{can the
effect on the ground state caused by adiabatic perturbation of $g$
be achieved through LOCC by Alice and Bob?} Differential local
convertibility of ground states is significant. Should LOCC
operations between Alice and Bob be capable of simulating a
particular physical process, then such a process is of limited
computational power, i.e., it is incapable of generating any quantum
coherence between $A$ and $B$.

\begin{figure}
\epsfig{file=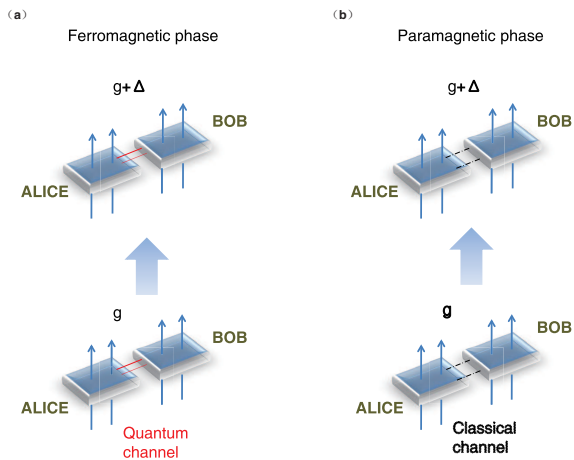,width=85mm}
 \caption{Differential local
convertibility illustration. Alice and Bob control the two
bi-partitions of a physical system whose ground state depends on a
Hamiltonian that can be varied via some external parameter $g$. In
one phase (panel a), the conversion from one ground state
$|G(g)\rangle$ to another $|G(g +\Delta)\rangle$ requires a quantum
channel between Alice and Bob, i.e, coherent interactions between
the two partitions are required. We say that this phase has no local
convertibility. After phase transition (panel b), Alice and Bob are
able to convert $|G(g)\rangle$ to $|G(g +\Delta)\rangle$ via only
local operations and classical communications, and local
convertibility becomes possible. If $\Delta\rightarrow0$, local
convertibility becomes differential local convertibility. This
implies that it is impossible to completely simulate the adiabatic
evolution of the ground state with respect to $g$ by LOCC in one
quantum phase phase and possible in the other phase.}
\label{fig1-sketch}
\end{figure}

We make use of the most powerful notion of differential local
convertibility, that of LOCC operations together with assisted
entanglement \cite{Nielsen, jonathanplenio}. Given some
infinitesimal $\Delta$, let $|G(g)\rangle_{AB}$ and
$|G(g+\Delta)\rangle_{AB}$ be the ground states of the given system
when the external parameter is set to be $g$ and $g + \Delta$
respectively. The necessary and sufficient conditions for local
conversion between $|G(g)\rangle_{AB}$ and
$|G(g+\Delta)\rangle_{AB}$ is given by $S_{\alpha}(g)\geq
S_{\alpha}(g +\Delta)$ for all $\alpha$, where
\begin{eqnarray}
S_{\alpha}(g)=\frac{1}{1-\alpha}\log_2[Tr\rho_A^{\alpha}(g)]=\frac{1}{1-\alpha}\log_2\big[\sum_{i=1}^d\lambda_i^{\alpha}\big]
\end{eqnarray}
is the R\'{e}nyi entropy with parameter $\alpha$, $\rho_A(g)$ is the
reduced density matrix of $|G(g)\rangle_{AB}$ with respect to
Alice's subsystem, and $\{\lambda_i\}$ are the eigenvalues of
$\rho_A(g)$ in decreasing
order\cite{necessarysufficient1,necessarysufficient2,necessarysufficient3}.
Thus, if the R\'{e}nyi entropies of two states intercept for some
$\alpha$, they cannot convert to each other by LOCC even in the
presence of ancillary entanglement\cite{ccf}. In the
$\Delta\rightarrow0^+$ limit, we may instead examine the sign of
$\partial_g S_{\alpha}(g)$ for all $\alpha$. If $\partial_g
S_{\alpha}(g)$ does not change sign, the effect of an infinitesimal
increase of $g$ results in global shift in $S_{\alpha}(g)$, with no
intersection between $S_{\alpha}(g+\Delta)$ and $S_{\alpha}(g)$.
Otherwise, an intersection must exist.

\section{Results}
\subsection{Transverse field Ising model.}
Before we consider the general $XY$ model, we highlight key ideas on
the transverse Ising model, which has the Hamiltonian
\begin{eqnarray}
H_I(g)=-\sum_{i=1}^N(\sigma_i^x\sigma_{i+1}^x+g\sigma_i^z),
\end{eqnarray}
where $\sigma^k$, for $k = x,y, z$ are the usual Pauli matrices and
periodic boundary conditions are assumed. The transverse Ising model
is one of the simplest models that has a phase transition, therefore
it often serves as a test bed for applying new ideas and methods to
quantum phase transitions. Osterloh {\it et al.} have previously
shown that the derivative of the concurrence is a indicator of the
phase transition\cite{Osterlohnature}. Nielsen {\it et al.} have
also studied concurrence between two spins at zero or finite
temperature\cite{nielsenIsing}. Recently, the Ising chain with
frustration has been realized in experiment\cite{L.M.Duan}.

The transverse Ising model features two different quantum phases,
separated by a critical point at $g=1$. When $g<1$, the system
resides in the ferromagnetic (symmetric) phase. It is ordered, with
nonzero order parameter $\langle\sigma^x\rangle$, that breaks the
phase flip symmetry $\Pi_i\sigma_i^z$. When $g>1$, the system
resides in the symmetric paramagnetic (symmetry broken) phase, such
that $\langle\sigma^x\rangle=0$.

There is systematic qualitative difference in the
computational power afforded by perturbation of $g$ within these two
differing phases. In the paramagnetic phase,
$\partial_g S_{\alpha}(g)$ is negative for all $\alpha$, hence
increasing the external magnetic field can be simulated by LOCC. In the ferromagnetic  $\partial_g
S_{\alpha}(g)$ changes signs for certain $\alpha$. Thus the ground
states are not locally convertible, and perturbing the magnetic
field in either direction results in fundamentally non-local quantum effects.

This result is afforded by the study of how R\'{e}nyi
entropy within the system behaves. From Eq.(1), we see that R\'{e}nyi entropy
contains all knowledge of $\{\lambda _i\}$\cite{Renyientropy}. For
large $\alpha$, $\lambda_i^\alpha$ vanishes when $\lambda_i$ is
small, and larger eigenvalues dominate. In the limit where
$\alpha\rightarrow\infty$, all but the largest eigenvalue
$\lambda_1$ may be neglected such that
$S_{\infty}=-\log_2\lambda_1$. In contrast, for small values of
$\alpha$, smaller eigenvalues become as important as their larger
counterparts. In the  $\alpha\rightarrow0^+$ limit, the R\'{e}nyi
entropy approaches to the logarithm of the rank for the reduced
density matrix, i.e, the number of non-zero eigenvalues.

This observation motivates  study of the eigenvalue spectrum. In
systems of finite size (Fig. 2 ($a$)), the largest eigenvalue
monotonically increases while the second monotonically decreases for
all $g$. All the other eigenvalues $\lambda_k$ exhibit a maximum at
some point $g_k$. From the scaling analysis (Fig. 2 ($b$) and
($c$)), we see that as we increase the size of the system, $g_k
\rightarrow 1$ for all $k$. Thus, in the thermodynamic limit,
$\lambda_k$ exhibits a maximum at the critical point of $g = 1$ for
all $k \geq 3$. Knowledge of this behavior gives intuition to our
claim.

\begin{figure}
\begin{center}
\epsfig{file=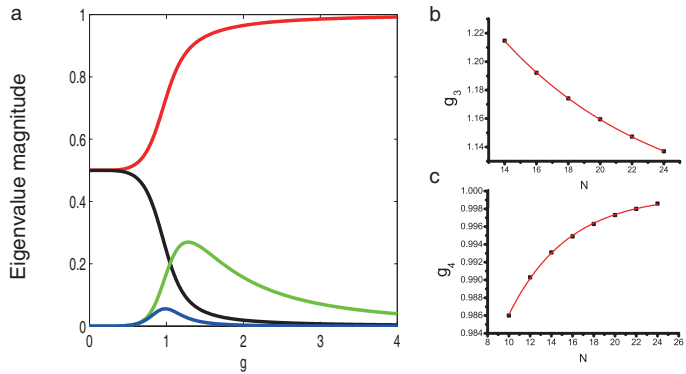,width=90mm}
\end{center}
\caption{The largest 4 eigenvalues and the scaling analysis. (a) The
four largest eigenvalues of transverse Ising ground state for N=10
case. The red line represents the largest eigenvalue $\lambda_1$,
and the black line is the second one $\lambda_2$. Note that we have
artificially magnified $\lambda_3$ (green) and $\lambda_4$ (blue) by
$10$ times for the sake of clarity since each subsequent eigenvalue
is approximately one order smaller than its predecessor. (b) The
scaling behaviors of the maximal points of the third eigenvalue.
When $N\rightarrow\infty$, the maximum points approach to the
critical point with certain acceptable error. The black dots are the
data and the red curves are the exponentially fit of these data. The
maximum point for the third eigenvalue
$g_3=0.495\times\exp(-\frac{N}{10.044})+1.09177$. (c) The maximum
point for the fourth eigenvalue
$g_4=-0.082\times\exp(-\frac{N}{5.527})+0.9996$. The maximums points
of smaller eigenvalues have similar behavior.}
\label{fig2-eigenvalues}
\end{figure}

In the ferromagnetic phase, i.e., $0<g<1$, $\partial_g
S_{\alpha}(g)$ takes on different signs for different $\alpha$. When
$\alpha \rightarrow0^+$, $S_{\alpha}$ tends to the logarithm of the
effective rank. From Fig. 2 we see that all but the two largest
eigenvalues increase with $g$, resulting in an increase of effective
rank. Thus $\partial_g S_{\alpha}(g) > 0$ for small $\alpha$. In
contrast, when $\alpha\rightarrow\infty$, $S_{\alpha} \rightarrow
-\log_2\lambda_1$. Since $\lambda_1$ increases with $g$ (Fig. 2),
$\partial_g S_{\alpha}(g) < 0$ for large $\alpha$. Therefore, there
is no differential local convertibility in the ordered phase.

In the paramagnetic phase, i.e., $g>1$, calculation yields that
$\partial_g S_{\alpha}(g)$ is negative for both limiting cases
considered above by similar reasoning. However, the intermediate
$\alpha$ between these two limits cannot be analyzed in a simple
way. The detail and formal proof of the result can be found in the
'Method' section, where it is shown that $\partial_g S_{\alpha}(g)$
still remains negative for all $\alpha>0$. Thus, differential local
convertibility exists in this phase.

These results indicate that at the critical point, there is
a distinct change in the nature of the ground state. Prior to the
critical point, a small perturbation of the external magnetic field
results in a change of the ground state that cannot be implemented
without two body quantum gates. In contrast, after phase transition,
any such perturbation may be simulated completely by LOCC.

\subsection{XY model.}
We generalize our analysis to the $XY$ model, with Hamiltonian
\begin{eqnarray}
H=-\sum_i[\frac{1}{2}(1+\gamma)\sigma_i^x\sigma_{i+1}^x+\frac{1}{2}(1-\gamma)\sigma_i^y\sigma_{i+1}^y+ g\sigma_i^z],
\end{eqnarray}
for different fixed values of $\gamma > 0$. The transverse Ising
model thus corresponds to the the special case for this general
class of models, in which $\gamma = 1$. For $\gamma \neq 1$, there
exists additional structure of interest in phase space beyond the
breaking of phase flip symmetry at $g = 1$. In particular, there
exists a circle, $g^2+\gamma^2 = 1$, on which the ground state is
fully separable. The functional form of ground state correlations
and entanglement are known differ substantially on either side of
the circle\cite{McCoy,Korepin,TZWei}, which motivates the
perspective that the circle is a boundary between two differing
phases. Indeed, such a division already exists from the perspective
of an entanglement phase diagram, where different `phases' are
characterized by the presence and absence of parallel entanglement
\cite{luigi}.

Analysis of local convertibility reveals that from the perspective
of computational power under adiabatic evolution, we may indeed
divide the system into three separate phases (Fig. 3). While the
disordered paramagnetic phase remains locally convertible, the local
convertibility of the ferromagnetic phase now depends on whether
$g^2+\gamma^2 > 1$. In particular, for each fixed $\gamma > 0$. The
system is only locally non-convertible when $g > \sqrt{1 -
\gamma^2}$. We summarize these results in a `local-convertibility
phase-diagram', where the ferromagnetic region is now divided into
components defined by their differential local convertibility.

\begin{figure}
\begin{center}
\epsfig{file=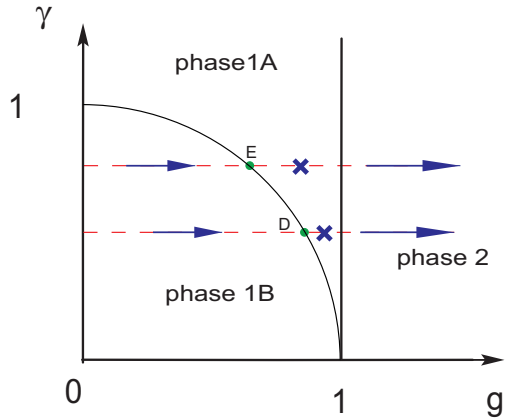,width=85mm}
\end{center}
\caption{ XY model local convertibility phase diagram. Consideration
of differential local convertibility separates the XY model in three
phases, which we label phase 1A, phase 1B and phase 2. We consider
differential local convertibility for fixed values of $\gamma$ while
$g$ is perturbed. Differential local convertibility is featured
within both phase 1B and phase 2, but not phase 1A.}
\label{fig3-phasediagram}
\end{figure}

\subsection{Different partitions.}
Numerical evidence strongly suggests that our results are not
limited to a particular choice of bipartition. We examine the
differential local convertibility when both systems of interest is
partitioned in numerous other ways, where the two parties may share
an unequal distribution of spins (Fig. 4). The qualitative
properties of $\partial_g S_{\alpha}$ remain unchanged. While it is
impractical to analyze all $2^N$ possible choices of bipartition,
these results motivate the conjecture that differential local
convertibility is independent of our choice of partitions. Should
this be true, it has strong implications: The computation power of
adiabatic evolution in different phases are drastic. In one phase,
perturbation of the external field can be completely simulated by
LOCC operations on individual spins, with no coherent two-spin
interactions. While, in other phases, any perturbation in the
external field creates coherent interactions between any chosen
bipartition of the system.

\begin{figure}
\begin{center}
\epsfig{file=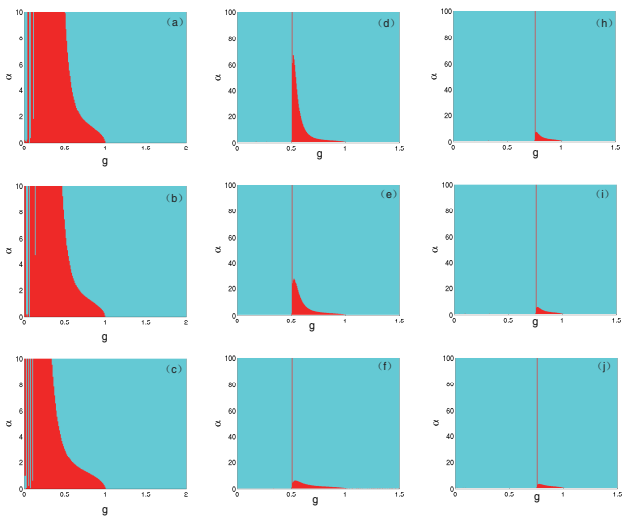,width=85mm}
\end{center}
\caption{The sign distribution of $\partial_g S_{\alpha}$ in Ising
model and XY model for different bi-partitions. The sign
distribution of $\partial_g S_{\alpha}$ on the $\alpha-g$ plane for
different bi-partitions on the systems. Panel $(a)$, $(b)$, $(c)$
correspond to Ising model of size $N=12$, where Alice possesses
$6,5,4$ of the spins respectively. $\partial_g S_{\alpha}$ is
negative in lighter regions and positive in the red regions.
Clearly, regardless of choice of bipartition, $\partial_g
S_{\alpha}$ is always negative for $g>1$ and takes on both negative
and positive values otherwise. Note that for very small $g$,
$\partial_g S_{\alpha}$ only becomes negative for very large
$\alpha$ and thus appears completely positive in the graph above.
The existence of negative $\partial_g S_{\alpha}$ can be verified by
analysis of $\partial_g S_{\alpha}$ in the $\alpha \rightarrow
\infty$ limit. The choice of bipartition affects only the shape of
the $\partial_g S_{\alpha}=0$ boundary, which is physically
unimportant. Panel $(d)$, $(e)$, $(f)$ correspond to XY model with
fixed $\gamma = \sqrt{3}/2$, $N=14$ and $L=7, 6, 5$ respectively.
Panel $(h)$, $(i)$, $(j)$ correspond to XY model with fixed $\gamma
= \sqrt{7}/4$, $N=18$ and $L=9,8,7$, respectively.  Here the value
of $L$ represents a bipartition in which $L$ qubits are placed in
one bipartition and $N-L$ qubits in the other. The only region in
which $\partial S_\alpha/\partial g$ takes on both negative and
positive values is in phase 1A of Figure 4. Note that the transition
between phase 1A and 1B occurs at $g=0.5$ for $\gamma = \sqrt{3}/2$
(Point E, in Fig 3) and $g=0.75$ for $\gamma = \sqrt{7}/4$ (Point D,
in Fig 3.)} \label{fig4-result}
\end{figure}

\section{Discussion}
The study of differential local convertibility of the ground state
gives direct operational significance to phase transitions in the
context of quantum information processing. For example, adiabatic
quantum computation (AQC) involves the adiabatic evolution of the
ground state of some Hamiltonian which features a parameter that
varies with time
\cite{Adiabaticquantumcomputation1,Adiabaticquantumcomputation2}.
This is instantly reminiscent of our study, which observes what
computational processes are required to simulate the adiabatic
evolution of the ground state under variance of an external
parameter in different quantum phases.

Specifically, AQC involves a system with Hamiltonian $(1-s) H_0 + s
H_p$, where the ground state of $H_0$ is simple to prepare, and the
ground state of $H_p$ solves a desired computational problem.
Computing the solution then involves a gradual increment of the
parameter $s$. By the adiabatic theorem, we arrive at our desired
solution provided the $s$ is varied slowly enough such that the
system remains in its ground state
\cite{Adiabaticquantumcomputation1,Adiabaticquantumcomputation2}. We can regard this process of computation from the perspective of
local convertibility and phase transitions. Should the system lie in
a phase where local convertibility exists, the increment of $s$ may
be simulated by LOCC. Thus AQC cannot have any computational
advantages over classical computation. Only in phases where no local
convertibility exists, can AQC have the potential to surpass
classical computation. Thus, a quantum phase transition could be
regarded as an indicator from which AQC becomes useful.

In fact, the spin system studied in this paper is directly relevant
to a specific AQC algorithm. The problem of ``$2-SAT$ on a Ring:
Agree and Disagree'' features an adiabatic evolution involving the
Hamiltonian
 \begin{eqnarray}
\widetilde{H}(s)=(1-s)\sum_{j=1}^N(1-\sigma_j^x)+s\sum_{j=1}^N\frac{1}{2}(1-\sigma_j^z\sigma_{j+1}^z),
 \end{eqnarray}
where $s$ is slowly varied from $0$ to $1$
\cite{Adiabaticquantumcomputation1,Adiabaticquantumcomputation2}.
This is merely a rescaled version of the Ising chain studied here,
where the phase transition occurs at $s=\frac{2}{3}$. According to
the analysis above, the AQC during the paramagnetic phase can be
simulated by local manipulations or classical computations. For the
period of ferromagnetic phase, we can do nothing to reduce the
adiabatic procedure.

In this paper, we have demonstrated that the computational power of adiabatic evolution in the $XY$ model is dependent on which quantum phase it resides in. This surprising relation suggests different quantum phases may not only have different physical properties, but may also display different computational properties. This hints that not only are the tools of quantum information useful as alternative signatures of quantum phase transitions, but that the study of quantum phase transitions may also offer additional insight into quantum information processing. This motivates the study of the quantum phases within artificial systems that correspond directly to well known adiabatic quantum algorithms, which may grant additional insight on how adiabatic computation relates to the physical properties of system that implements the said computation. There is much potential insight to be gained in applying the methods of analysis presented here to more complex physical systems that featuring more complex quantum transitions.

In addition, differential local convertibility also may possess
significance beyond information processing. One of the proposed
indicators of a topological order involves coherent interaction
between subsystems that scale with the size of the
system\cite{wenbook,Wen}. In our picture, such a indicator could
translate to the requirement for non-LOCC operations within
appropriate chosen bipartite systems. Thus, differential local
convertibility may serve as an additional tool for the analysis of
such order.

\section{Methods}

\subsection{Eigenvalue properties.}
For the transverse field Ising model, the largest eigenvalue
$\lambda_1$ monotonically increases while the second $\lambda_2$
monotonically decreases for all $g$ (Fig. 2). In the thermodynamic
limit all the other eigenvalues increase in the $g<1$ region and
decrease in the $g>1$ region. Moreover, the eigenvalues other than
the largest two are much smaller than $\lambda_1$ and $\lambda_2$.
Therefore we can average them when considering their contribution to
the Renyi entropy. Thus the eigenvalues are assumed to be
$0.5+\delta, 0.5-\epsilon, (\epsilon-\delta)/(2^n-2),
(\epsilon-\delta)/(2^n-2), \dots$ when $g<1$, and
$1-\delta^{\prime}-\epsilon^{\prime}, \epsilon^{\prime},
\delta^{\prime}/(2^n-2), \dots$, when $g>1$, where $n$ is the
particle number belonging to Alice and certainly Bob has the other
$N-n$ particles. Because of some obvious reasons such as
$\lambda_1>\lambda_2>\lambda_3 \cdots$ and all these eigenvalues are
positive, and so on, we can derive the following relations easily:
$0<\delta<\epsilon<0.5, 0<\partial\delta/\partial
g<\partial\epsilon/\partial g,
0<\delta^{\prime}<\epsilon^{\prime}<0.5,
\partial\delta^{\prime}/\partial g<0$, and
$\partial\epsilon^{\prime}/\partial g<0$. Then we can prove the main
result for each phase region. Namely, in the $g<1$ phase,
$\partial_g S_{\alpha}$ is positive for small $\alpha$ but negative
for large $\alpha$; and for the $g>1$ phase, $\partial_g
S_{\alpha}<0$ for all $\alpha$.

\subsection{Ferromagnetic phase.}
In the ferromagnetic phase $g<1$, the eigenvalues are $0.5+\delta,
0.5-\epsilon, (\epsilon-\delta)/(2^n-2), (\epsilon-\delta)/(2^n-2),
\dots$. So the Renyi entropy
\begin{small}
\begin{equation}
S_{\alpha}=\frac{1}{1-\alpha}\log\Big[(0.5+\delta)^{\alpha}+(0.5-\epsilon)^{\alpha}
+(2^n-2)\big(\frac{\epsilon-\delta}{2^n-2}\big)^{\alpha}\Big],
\end{equation}
and
\begin{flalign}
\begin{split}
\frac{\partial S_{\alpha}}{\partial
g}=\frac{1}{1-\alpha}\frac{1}{(0.5+\delta)^{\alpha}+(0.5-\epsilon)^{\alpha}+(2^n-2)\big(\frac{\epsilon-\delta}{2^n-2}\big)^{\alpha}}\nonumber\\
\end{split}&
\end{flalign}
\begin{flalign}
\begin{split}
\times\alpha\Big[(0.5+\delta)^{\alpha-1}\frac{\partial\delta}{\partial
g}-(0.5-\epsilon)^{\alpha-1}\frac{\partial\epsilon}{\partial g}
+\big(\frac{\epsilon-\delta}{2^n-2}\big)^{\alpha-1}\big(\frac{\partial\epsilon}{\partial
g}-\frac{\partial\delta}{\partial g}\big)\Big]\nonumber\\
\end{split}&
\end{flalign}
\begin{flalign}
\begin{split}
\qquad
=\frac{\alpha}{1-\alpha}\frac{1}{(0.5+\delta)^{\alpha}+(0.5-\epsilon)^{\alpha}+(2^n-2)\big(\frac{\epsilon-\delta}{2^n-2}\big)^{\alpha}}\nonumber\\
\end{split}&
\end{flalign}

\begin{flalign}
\times\Big\{\frac{\partial\delta}{\partial
g}[(0.5+\delta)^{\alpha-1}-\big(\frac{\epsilon-\delta}{2^n-2}\big)^{\alpha-1}]-\nonumber\\
\frac{\partial\epsilon}{\partial
g}[(0.5-\epsilon)^{\alpha-1}-\big(\frac{\epsilon-\delta}{2^n-2}\big)^{\alpha-1}]\Big\}.
\end{flalign}
\end{small}

In the thermodynamic limit $N\rightarrow \infty$,
$(\epsilon-\delta)/(2^n-2)\rightarrow$ $0$.

When $0<\alpha<1$,
$\big(\frac{\epsilon-\delta}{2^n-2}\big)^{\alpha-1}\rightarrow
\infty$,
\begin{small}
\begin{eqnarray}
\frac{\partial S_{\alpha}}{\partial
g}=\frac{1}{1-\alpha}\frac{\alpha}{\epsilon-\delta}\big(\frac{\partial\epsilon}{\partial
g}-\frac{\partial\delta}{\partial g}\big)>0.
\end{eqnarray}
\end{small}

When $\alpha>1$,
$\big(\frac{\epsilon-\delta}{2^n-2}\big)^{\alpha-1}\rightarrow 0$,
\begin{small}
\begin{eqnarray}
\frac{\partial S_{\alpha}}{\partial
g}\sim\frac{1}{1-\alpha}[\frac{\partial\delta}{\partial
g}\big(1+\frac{\epsilon+\delta}{0.5-\epsilon}\big)^{\alpha-1}-\frac{\partial\epsilon}{\partial
g}].
\end{eqnarray}
\end{small}
As $\big(1+\frac{\epsilon+\delta}{0.5-\epsilon}\big)^{\alpha-1}>1$,
and $0<\partial\delta/\partial g< \partial\epsilon/ \partial g $, we
can see that the solution of $\partial S_{\alpha} / \partial g=0$
(labeled as $\alpha_0$) always exists in the region
$g<1\bigcap\alpha>1$, and $\partial S_{\alpha}/\partial g$ will be
negative as long as $\alpha>\alpha_0$. Moreover we can also see that
the smaller $g$ is, the smaller $\delta$ and $\epsilon$ are, and the
smaller $\big(1+(\epsilon+\delta)/(0.5-\epsilon)\big)$ is, and
therefore the larger $\alpha_0$ should be. This explains why we need
to examine larger value of $\alpha$ to find the crossing when $g$ is
very small.

Notice that in the above analysis we used the $N\rightarrow\infty$
condition in the $g<1$ region. We can also find that in Fig. 4 for
finite $N=12$ there is some small green area in the region
$\alpha<1\bigcap g<1$. However, in the above analysis of infinite
$N$, this area should be totally red. This difference is due to the
finite size effect.

\subsection{Paramagnetic phase.}
In the paramagnetic phase, $g>1$. The eigenvalues are
$1-\delta^{\prime}-\epsilon^{\prime}, \epsilon^{\prime},
\delta^{\prime}/(2^n-2), \dots$. The R\'{e}nyi entropy
\begin{small}
\begin{eqnarray}
S_{\alpha}=\frac{1}{1-\alpha}\log[(1-\delta^{\prime}-\epsilon^{\prime})^{\alpha}+(\epsilon^{\prime})^{\alpha}+(2^n-2)\big(\frac{\delta^{\prime}}{2^n-2}\big)^{\alpha}].
\end{eqnarray}
\end{small}
And
\begin{small}
\begin{flalign}
\begin{split}
\frac{\partial S_{\alpha}}{\partial
g} =  \frac{1}{1-\alpha}\frac{1}{(1-\delta^{\prime}-\epsilon^{\prime})^{\alpha}+(\epsilon^{\prime})^{\alpha}+(2^n-2)\big(\frac{\delta^{\prime}}{2^n-2}\big)^{\alpha}}\nonumber\\
\end{split}&
\end{flalign}
\begin{align}
\begin{split}
\times\alpha\Big[-(1-\epsilon^{\prime}-\delta^{\prime})^{\alpha-1}\big(\frac{\partial\delta^{\prime}}{\partial
g}+\frac{\partial\epsilon^{\prime}}{\partial g}\big)\nonumber\\
\end{split}&
\end{align}
\begin{align}
\begin{split}+(\epsilon^{\prime})^{\alpha-1}\frac{\partial\epsilon^{\prime}}{\partial
g}+\big(\frac{\delta^{\prime}}{2^n-2}\big)^{\alpha-1}\frac{\partial\delta^{\prime}}{\partial
g}\Big]\nonumber\\
\end{split}&
\end{align}
\begin{flalign}
\begin{split}
\qquad
 =\frac{\alpha}{1-\alpha}\frac{(1-\delta^{\prime}-\epsilon^{\prime})^{\alpha-1}}{(1-\delta^{\prime}-\epsilon^{\prime})^{\alpha}+(\epsilon^{\prime})^{\alpha}+(2^n-2)\big(\frac{\delta^{\prime}}{2^n-2}\big)^{\alpha}}\nonumber\\
 \end{split}&
\end{flalign}
\begin{flalign}
\begin{split}
\qquad \quad \times\Big\{\frac{\partial\delta^{\prime}}{\partial
g}\Big[\big(\frac{1}{2^n-2}\frac{\delta^{\prime}}{1-\delta^{\prime}-\epsilon^{\prime}}\big)^{\alpha-1}-1\Big]\nonumber\\
\end{split}&
\end{flalign}
\begin{align}
\begin{split}
+\frac{\partial\epsilon^{\prime}}{\partial
g}\Big[\big(\frac{\epsilon^{\prime}}{1-\epsilon^{\prime}-\delta^{\prime}}\big)^{\alpha-1}-1\Big]\Big\},
\end{split}&
\end{align}
\end{small}
where $\partial\delta^{\prime}/\partial
g,\partial\epsilon^{\prime}/\partial g<0$; and
$\epsilon^{\prime}/(1-\epsilon^{\prime}-\delta^{\prime}),\delta^{\prime}/(1-\epsilon^{\prime}-\delta^{\prime})(2^n-2)\in(0,1)$,
since $\lambda_1>\lambda_2>\lambda_3$.

So when $\alpha>1$, $[(\dots)^{\alpha-1}-1]<0$, $\{\dots\}>0$,
$\alpha/(1-\alpha)<0$. We have $\partial S_{\alpha}/\partial g<0$;
and when $0<\alpha<1$, $[(\dots)^{\alpha-1}-1]>0$, $\{\dots\}<0$,
$\alpha/(1-\alpha)>0$. We also have $\partial S_{\alpha}/\partial
g<0$. Hence, we can obtain that when $g>1$, $\partial
S_{\alpha}/\partial g<0$ for all $\alpha>0$. If we consider the full
condition for the local convertibility which includes the
generalization of $\alpha$ to negative value, it also can be proved
easily that the local convertible condition $\partial_g
S_{\alpha}/\alpha>0$ for all $\alpha$ is satisfied in the $g>1$
phase. As for the $g<1$ phase, the sign changing in the positive
$\alpha$ already violates the local convertible condition, so that
we do not need to consider the negative $\alpha$ part. In fact,
generally speaking, for the study of \textit{differential local
convertibility}, we can only focus on the positive $\alpha$ part,
because the derivative of R\'{e}nyi entropy over the phase
transition parameter will necessarily generate a common factor
$\alpha$, which will cancel the same $\alpha$ in the denominator.

To conclude, In the $g>1$ phase $\partial S_{\alpha}/\partial g$ is
negative for all $\alpha$, and in the $g<1\bigcap\alpha<1$ region it
is positive, while in the $g<1\bigcap\alpha>1$ region it can be
either negative or positive with the boundary depending on the
solution of Eq.(4).

Acknowledgement--- The authors thank Z. Xu, W. Son and L. Amico for
helpful discussions. J. Cui and H. Fan thank NSFC grant (11175248)
and ``973'' program (2010CB922904) for financial support. J. Cui, M.
Gu, L. C. Kwek, M.F. Santos and V. Vedral thank the National
Research Foundation \& Ministry of Education, Singapore for
financial support.

Author Contributions--- J.C. carried out the numerical work and
calculation. H.F., J.C., L.C.K. and M.G. contributed to the
development of all the pictures. J.C., M.G. and L.C.K. drafted the
paper. All the authors conceived the research, discussed the
results, commented on and wrote up the manuscript.

Competing Interests--- The authors declare that they have no
competing financial interests.

Correspondence--- Correspondence and requests for materials should
be addressed to J.C.

\end{document}